# Simulation of Urban Expansion and Farmland Loss in China by Integrating Cellular Automata and Random Forest

Yao Yao, Xiaoping Liu, Dachuan Zhang, Zhaotang Liang, Yatao Zhang

Y. Yao, X. Liu, D. Zhang

School of Geography and Planning, Guangdong Key Laboratory for Urbanization and Geo-simulation, Sun Yat-sen University, Guangzhou 510275, Guangdong province, China.

Z. Liang • Y. Zhang

School of Geography and Planning, Guangdong Key Laboratory for Urbanization and Geo-simulation, Sun Yat-sen University, Guangzhou 510275, Guangdong province, China.

**Abstract:** China has encountered serious land loss problems along with urban expansion due to rapid urbanization. Without considering complicated spatiotemporal heterogeneity, previous studies could not extract urban transition rules at large scale well. This study proposed a random forest algorithm (RFA) based cellular automata (CA) model to simulate China's urban expansion and farmland loss in a fine scale from 2000 to 2030. The objectives of this study are to 1) mine urban conversion rules in different homogeneous economic development regions, and 2) simulate China's urban expansion process and farmland loss at high spatial resolution (30 meters). Firstly, we clustered several homogeneous economic development regions among China according to official statistical data. Secondly, we constructed a RFA-based CA model to mine complex urban conversion rules and carried out simulation of urban expansion and farmland loss at each homo-region. The proposed model was implemented on Tianhe-1 supercomputer located in Guangzhou, China. The accuracy evaluation demonstrates that the simulation result of proposed RFA-based CA model is more in agreement with actual land use change. This study proves that the primary factor of farmland loss in China is rapid urbanization from 2000, and the farmland loss rate is expected to slow down gradually and will stabilize from 2010 to 2030. It shows that China is able to preserve the 1.20 million km farmland without crossing the "red line" within the next 20 years, but the situation remains severe.

**Highlights:**

- A large-scale geographical simulation framework via RFA-based CA is proposed.

- Driving forces of urban development at each homo-regions are extracted.

- The primary factor of farmland loss is the rapid urbanization of China since 2010.

- Farmland loss rate of China is expected to slow down gradually and will stabilize.

# 1 Introduction

Human-induced land-cover change, especially urban expansion, is a major threat to natural and ecological environment (Liu *et al.* 2014, Seto *et al.* 2012, Seto *et al.* 2012). As the most populous country in the world, China is now in a rapid urbanization process and causes many serious problems like environmental pollution and farmland loss (Chen *et al.* 2010, Ding 2003, Jiang *et al.* 2013, Liu *et al.* 2014, Lu *et al.* 2015, Tan *et al.* 2005). According to the National Bureau of Statistics, the farmland areas of China have sharply decreased from 1.28 million km$^2$ in 2000 to 1.22 million km$^2$ in 2010 as a result of rapid urbanization. To achieve sustainable development, the Chinese government formulated the "arable-land red line" policy to retain no less than 1.20 million km$^2$ (1.8 billion mu) farmland in 2003(Ding 2003, Liu *et al.* 2014, Xinhuanet 2012, Xinhuanet 2013). Therefore, precise simulation of urban expansion is of great importance to strictly control urban land uses and help develop countermeasures of farmland loss. .

Urban expansion and farmland loss are both very complex geographical processes. Over the past two decades, cellular automata (CA) have been widely and successfully applied to the simulation of complex geographical phenomena and lots of efforts were made for urban transition rule mining (Chen *et al.* 2008, Cheng and Masser 2002, Li *et al.* 2011, Li *et al.* 2013, Liu *et al.* 2008, Liu *et al.* 2014, White and Engelen 1993). Recently, many studies have integrated machine learning (ML) algorithms with CA models to mine urban transition rule, including logistical regression, ant colony optimization (ACO), immune algorithms and neural networks. (Arsanjani *et al.* 2013, Chen *et al.* 2008, Feng *et al.* 2011, Huang and Gao 2011, Li *et al.* 2013, Li *et al.* 2013, Li and Yeh 2002, Lin and Li 2015, Liu *et al.* 2008, Liu *et al.* 2010,

Liu *et al.* 2012, Shao *et al.* 2015, Yang *et al.* 2012). However, there are some shortcomings of these ML algorithms for transition rule mining. Logistic regression need the input variables to be linearly independent (Knol *et al.* 2012), while generally most spatial variables inputted in logistic-based CA model have certain relationship, i.e., cells that are close to the urban center are prone to be close to roads; ant colony optimization and immune algorithm have strong adaptability and parameter optimization ability, but the total computation efficiency is low and easy to fall into the local optimal ; neural network can address the above problem, but the internal mechanism of training process is still not clear and easy to be overfitting, and it is also time-consuming and hard to parallelize (Hung and Hung 2014). And most of the ML-CA models mentioned above were mainly conducted at small scale like a single city. As we known, urban expansion is a very complex nonlinear system, which is not only affected by the well-known factors like the distance to downtown and the terrain, but also the uncertainties like governmental policy, population migration and natural disasters (Liu *et al.* 2012). Therefore, in order to equip CA for large-scale nationwide simulation, we need to develop a nonlinear method which is high-precision, hard to be over-fitted, easy to parallelize and high-tolerance for random variables for mining urban expansion rules.

Most of the existing CA applications are at city-scale which aim at simulating the urban phenomena(He *et al.* 2015, Liu *et al.* 2007, Liu *et al.* 2010, Luo and Wei 2009, Shao *et al.* 2015, Wu *et al.* 2010, Xie *et al.* 2012, Yang *et al.* 2012). Li and Yeh indicated that the cities with large-scale area have diverse patterns and different processes of land-use changes (Li and Gar-On Yeh 2004), so it is unreasonable to only apply a single transition rule for large-scale simulation at national level. Some attempts have been made to enhance CA model's ability for large-scale simulation. For example, some studies have carried out provincial simulation of urban expansion by using case-based reasoning (CBR) and formal concept analysis (FCA) coupled with high performance computing technology (Li and Liu 2006, Lin and Li 2015). Though high performance computing technology was already used to improve the computational efficiency of large-scale CA simulations in this case, however, its performance was still unsatisfactory for detailed urban simulation (Li *et al.* 2010, Li *et al.* 2012). The aforementioned parallel CA models have parallel-unfriendly algorithm for transition rules mining and the hardware platforms they used limited the parallel computing performance. (Li *et al.* 2012) Thus, these models are not suitable for detailed geographical simulations in large scale (Liu *et al.* 2014). In this study, we propose a simulation model which is suitable for large-scale detailed simulation and high-performance computing to address these issues.

In this paper, we designed a RFA-based CA model to simulate China's urban expansion and farmland loss process at resolution of 30 meters. Firstly, we distributed the entire country into different homo-regions based on the similarity of socio-economic conditions at the provincial level. Then we introduced the random forest algorithm (RFA) to mine China's urban conversion rules in different regions as well as to simulate China's urban expansion and farmland loss from 2000 to 2030. The RFA-based CA model also measured the importance of each spatial variable according to its contribution for predicting urban transition probability during the training process. Therefore, the simulation results can offer us explanation about the driving forces of rapid urbanization in China. The proposed RFA-CA model was implemented on Tianhe-1 supercomputer and the ANN-based CA model was also conducted on the same computing environment to demonstrate the robustness and effectiveness of our proposed model..

## 2 Study area and data

## 2.1 Study area

Our study case was conducted in the whole territory of China (Figure 1), which is about 9,600,000 km$^2$ and contains 23 provinces, 5 autonomous regions, 4 direct-controlled municipalities (Beijing, Tianjin, Shanghai, and Chongqing), and 2 mostly self-governing special administrative regions (Hong Kong and Macau). China is such one country that feeds approximate one-fifth of the human population by less than one-tenth of the earth's cultivated land. A considerable urban expansion has witnessed since 1978 because of the fast economic development in this area, which has brought rapid loss of farmland resource (Zhang *et al.* 2012).

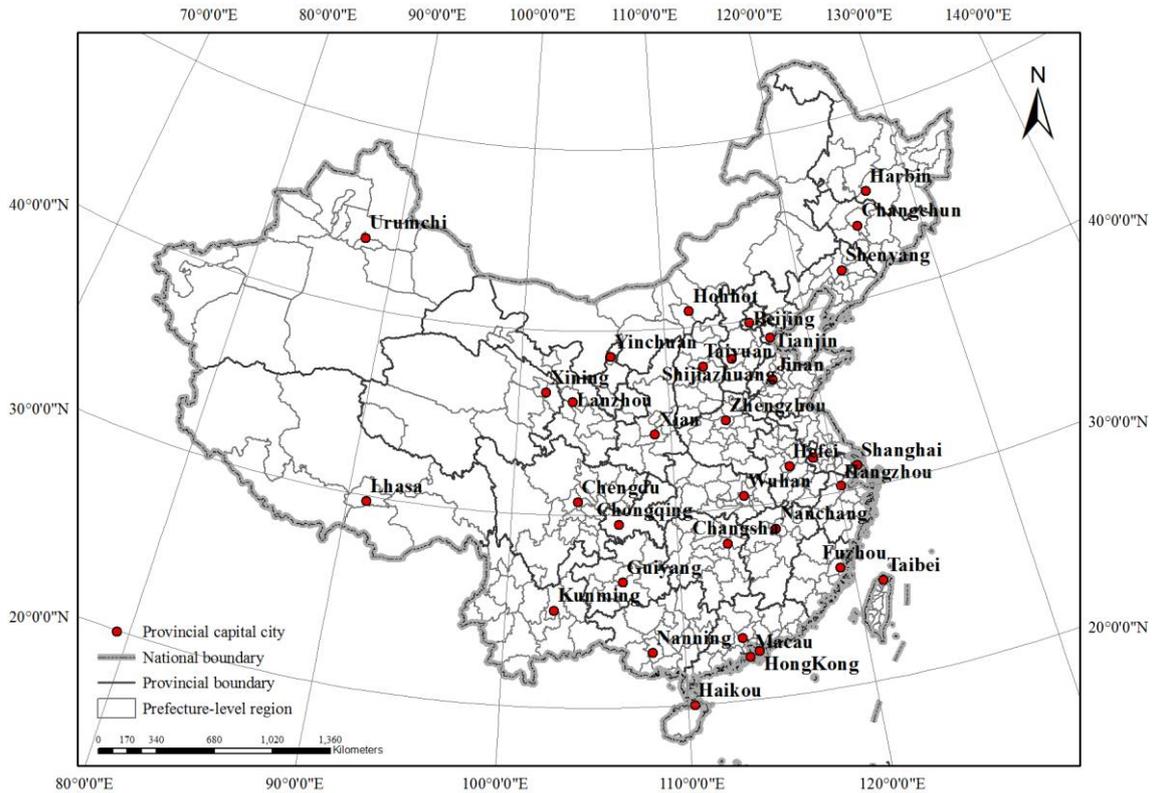

**Figure 1.** The study area of China with 34 provincial-level administrator regions

## 2.2 Data acquisition and pre-processing

The vector dataset of China's administrator border (1:4,000,000), which contains national level and provincial level boundaries, was obtained from DIVA-GIS (http://www.diva-gis.org/gdata). The border dataset contains 34 polygon features and covers all the provinces and province-level municipalities in China. Through National Data website of National Bureau of Statistics of China (http://data.stats.gov.cn/), we obtained every province's economy development data, including GDP, growth rate of GDP, Per Capita GDP (GDPPC), growth rate of GDPPC, proportion of primary industry, proportion of secondary industry, proportion of tertiary industry, permanent resident population, total population, etc. Also, the Annual Official Journal of China's Ministry of Land and Resources has released a large amount of statistical data from 1990 to 2014 (http://www.mlr.gov.cn/zwgk/tjxx/), including the area and types of farmland of each province. In this study, we used data from these two resources to construct economic development database for large-scale simulation of urban expansion and farmland loss.

Land cover data is the most important data in this study. Figure 2 shows China's Land cover from the classifications of Landsat images in 2000 and 2010 at resolution of 30 meters (Chen *et al.* 2015). The land cover data was obtained from Chinese Academy of Surveying and Mapping (CASM). According to the land cover classification standards of CASM, land cover data are divided into eleven categories (Farmland, forest, grassland, shrub land, wetland, waterbodies, tundra, artificial surfaces, bare land, permanent snow and ice, unknown place). In order to simplify land use types and better describe the conversion relation between urban expansion and farmland loss, we have reclassified these categories into three custom categories as urban, non-urban and limited-development area and thus generated the "actual land cover" datasets (Urban area: artificial surfaces; non-urban area: farmland, forest, grassland, shrub land and bare land; limited-development area: wetland, waterbody, tundra and permanent snow/ice). Also nine types of land use conversion relation were generated.

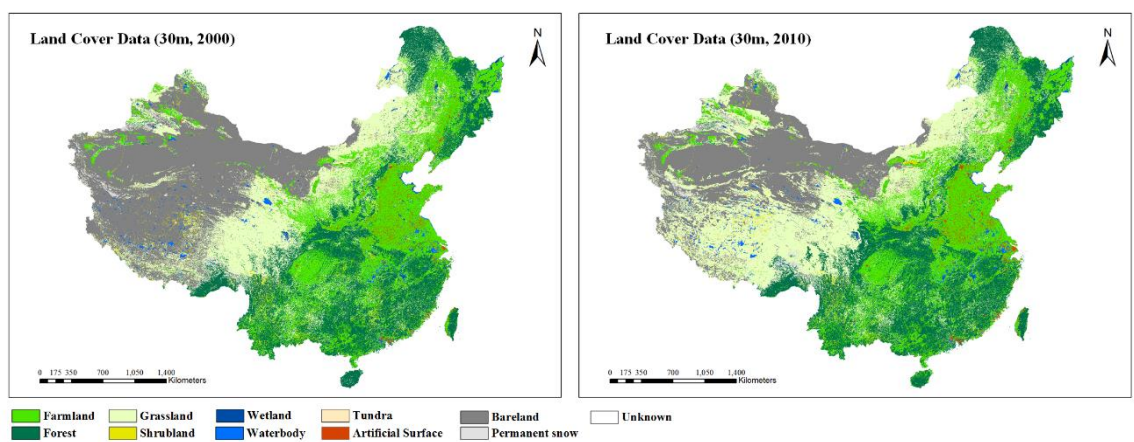

**Figure 2.** China's Land cover from the classifications of Landsat images in 2000 and 2010 (Data source: Chinese Academy of Surveying and Mapping).

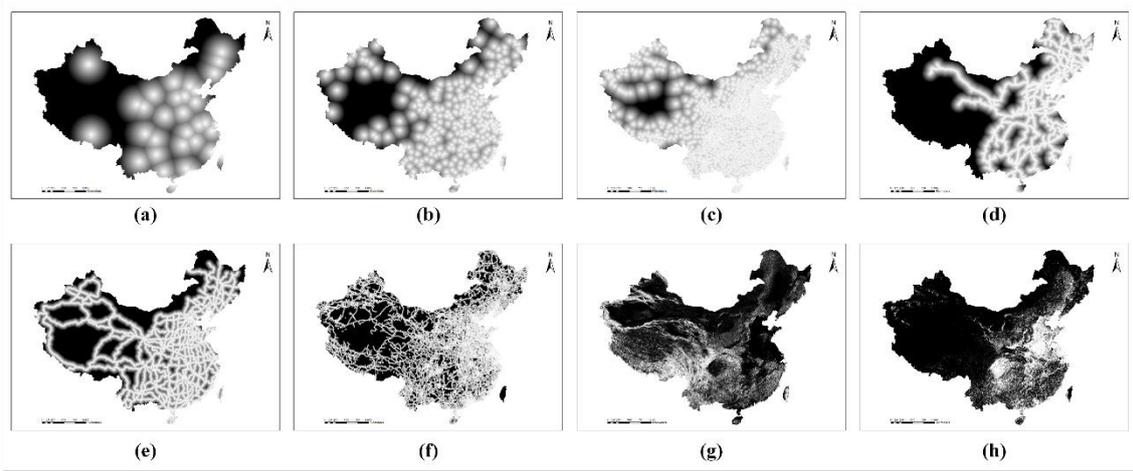

**Figure 3** Auxiliary spatial variables in this study: (a) distance to provincial capital cities, (b) distance to prefecture-level cities, (c) distance to counties, (d) distance to main railways, (e) distance to main roads, (f) distance to other roads, (g) DEM slope (30 meters), (h) coarse population dataset (LandScan)

Several auxiliary spatial data sets were also applied in this study. The distance to the road network and important cities are related directly to the level of urbanization, especially in developing countries (Poston Jr and Yaukey 2013). Besides, the population is related closely to the level of urbanization, which affects the intensity of land development a lot. Figure 3 shows the auxiliary spatial variables. The spatial distance data and DEM slope data were derived from basic GIS datasets from CASM, and the population data of China was derived from global population dataset of The LandScan Global Population Project (http://web.ornl.gov/sci/landscan/).

In this study, we developed a piecewise normalization method for spatial data pre-processing. As shown below, the mean $\mu$ and standard deviation $\sigma$ of the spatial data $X$ were calculated before value $X$ were normalized into [0, 1] by employing three-sigma rule (Grafarend 2006) (Equations (1) to (4)).

$$\mu = \frac{\sum_{i=1}^{N} x_i}{N} \tag{1}$$

$$\sigma = \frac{\sum_{i=1}^{N}\sqrt{(\mu-x_i)^2}}{N} \quad (2)$$

$$\begin{cases} X_1 = \mu - 3\sigma > \min? \mu - 3\sigma : \min \\ X_2 = \mu + 3\sigma < \min? \mu - 3\sigma : \max \end{cases} \quad (3)$$

$$x_i' = \begin{cases} 0 & x_i < X_1 \\ (x_i - X_1)/(X_2 - X_1) & X_2 \leq x_i \leq X_1 \\ 1 & x_i > X_2 \end{cases} \quad (4)$$

Where $\mu$ and $\sigma$ are mean and standard deviation respectively. $x_i$ is the $i$-th spatial variable, $N$ is the total number of spatial variables, and $\max$ and $\min$ denote the maximum and minimum of spatial dataset $X$.

# 3 Methodology

The purpose of our study is to extract the nationwide urban conversion rules by using random forest algorithm (RFA) integrated CA model and to simulate urban expansion and its impacts on the farmland loss. In this study, we took three steps to simulate the farmland loss in every homo-region: 1) We clustered provincial-level administrator regions of China into several homogeneous regions according to official statistical data of economic development status; 2) We developed RFA-based CA model to compute the overall urban conversion probability and extract the urban conversion rules in each region; 3) We simulated urban expansion and farmland loss based on the urban conversation rules extracted by proposed RFA-based CA model, and measured the reliability of our results by comparing with results from ANN-based CA models.

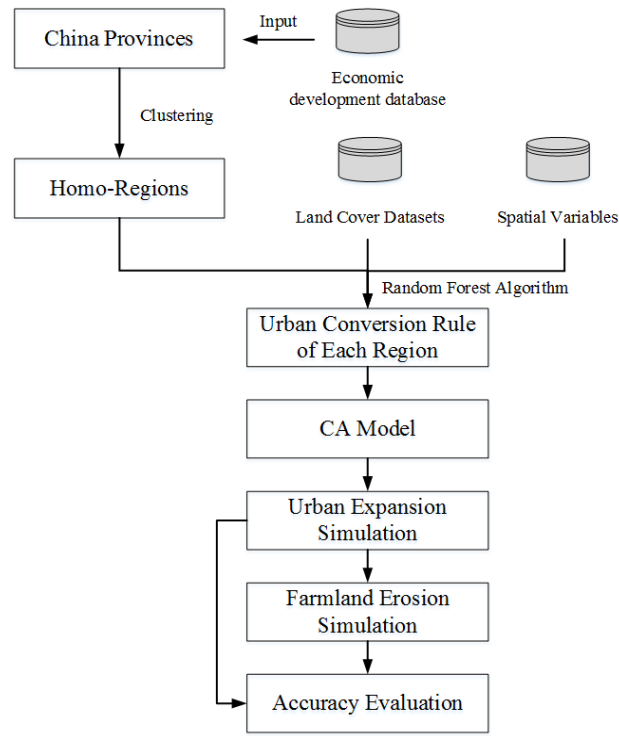

**Figure 4** Flow chart of urban expansion and farmland loss simulation by using RFA-based CA models

## 3.1 Regional agglomeration

Considering China's great differences in geographical resource distribution and complicated socio-economic development status, driving forces of land development are quite varied in different regions of China.    Besides, due to multi-scale effect, the change in spatial distribution is a course of gradual accumulation and one from quantitative to qualitative (LAND *et al.* 1999, Xu *et al.* 2001). Therefore, by taking multi-scale effect account, our study performed regional agglomeration method in units of provinces and provincial level municipalities in order to cluster the research area into distinct homogeneous regions for urban conversion rules extraction. Through the regional agglomeration, we could obtain more suitable homogeneous socio-economic regions for rules-mining; also the parallel computing could be applied to these clusters and thus reduced computation time and balanced computation resources.

Firstly, the socio-economic indexes of each province were normalized into range [0, 1] by minimum-maximum normalization method, including growth rate of GDP, per Capita GDP(GDPPC) and growth rate of GDPPC, etc. Then principal component analysis was performed on the normalization results. In the next step, the first two components of PCA result were put into SPSS to carry out cluster analysis using the centroid clustering method and the Pearson correlation for metrics. At last, the non-adjacent homogeneous regions were divided into different groups based on socio-economic indexes.

## 3.2 CA Model Integrating RFA

In this section, we will first introduce the method of Random Forest Algorithm (RFA) and then deploy it to construct our urban development model. The mathematical foundation of Random Forest Algorithm can be found in the work of Breiman (Breiman 2001).

### 3.2.1 Random Forest Algorithms (RFA)

RFA is a multi-classifier combination model that comprises a large number of decision trees $\{h(X, \Theta_k), k = 1, ..., M\}$. Specifically, the parameter set $\Theta_k$ of each decision tree consists of mutual independent random variables with the same distribution. RFA assumes the existence of $N$ classification categories $\{Y_i, i = 1, 2...N\}$ and randomly draws $M$ samples from the original training dataset to build $M$ decision tree classifiers. The size of each random sample is the same as that of the original training dataset. Each decision tree classifier will yield a classification result when the independent variable $X$ is given, which stands for one vote for the final classification result. Therefore, $M$ classification results will be separately obtained after the unclassified data $\theta$ being inputted to $M$ decision tree classifiers. By taking each tree's classification result as one vote, we can calculate the probabilities $\{P_i, i=1,2...N\}$ of unclassified data $\theta$ falling into each category $\{Y_i, i=1,2...N\}$, where $\sum_{i=1}^{N} P_i = 1$.

By selecting the features of each sample, RFA can generate different training sub-datasets to increase the diversity between classifiers, so as to minimize the potential for over-fitting during the training process and improve predictive ability. RFA is a multi-classifier combination system generated by training decision tree classifiers $M$. Equation (5) measures the probability for unclassified data $\theta$ falling into each category and Equation (6) generates the final classification result (Biau 2012, Breiman 2001).

$$P_i(x) = \frac{I(h_i(x) = Y_i)}{M} \tag{5}$$

$$H(x) = \arg\max_{Y_i} \sum_{i=1}^{M} I(h_i(x) = Y_i) \tag{6}$$

Specifically, $H(x)$ is the result of the multi-classifier combination model, $h_i(x)$ is the result of the single decision tree, $Y_i$ is the classification result of the single tree, and $I(\bullet)$ denotes the indicator function. Equations (5) and (6) illustrate that the classification result of RFA is based on a majority voting rule.

Furthermore, RFA is more than an assembly of decision tree classifiers. Previous study has indicated that 37% of the samples in original dataset $D$ will not be extracted and will be classified as out-of-bag (OOB) data (Biau 2012). Using OOB data to estimate the performance and accuracy of the RFA classification model is referred to as OOB-estimation. We can obtain an error report of OOB-estimation for each decision tree. The generalization error of RFA can be calculated by averaging the errors of the decision trees via OOB-estimation.

In our study, we implemented RFA classifier by using the Alglib package

(http://www.alglib.net/ ).In order to obtain the contribution of each type of spatial variable, the random noises are added into the spatial variables of $i-th$ type during the training process ($i \in [1,V]$, $V$ is the types count of spatial variables), and thus we can obtain the error of RFA classifier $DF_{error}^{i}$. Assumed that $AE_i$ is the predict average error of original training dataset by using $i-th$ error RFA classifier $DF_{error}^{i}$, and $AE_{true}$ is the predicted average error of original training dataset by using the correct RFA classifier $DF_{correct}$, then the contribution value of $i-th$ type of spatial variables can be obtained by Equations (7) as follows(Fakhraei *et al.* 2014, Palczewska *et al.* 2014).

$$Contribution_i = \frac{abs(AE_i - AE_{true})}{\sum_{i=1}^{V} abs(AE_i - AE_{true})} \quad (7)$$

### 3.2.2 RFA-Based CA model

RFA integrated CA model is developed to simulate and predict urban expansion and farmland loss at large-scale in this section. As shown in Figure 5, our RFA-based CA model involves two parts: training and simulating processes. The objective of training process is to mine the urban conversion probability of each region, and the simulating process is to predict the urban expansion and farmland loss of each region.

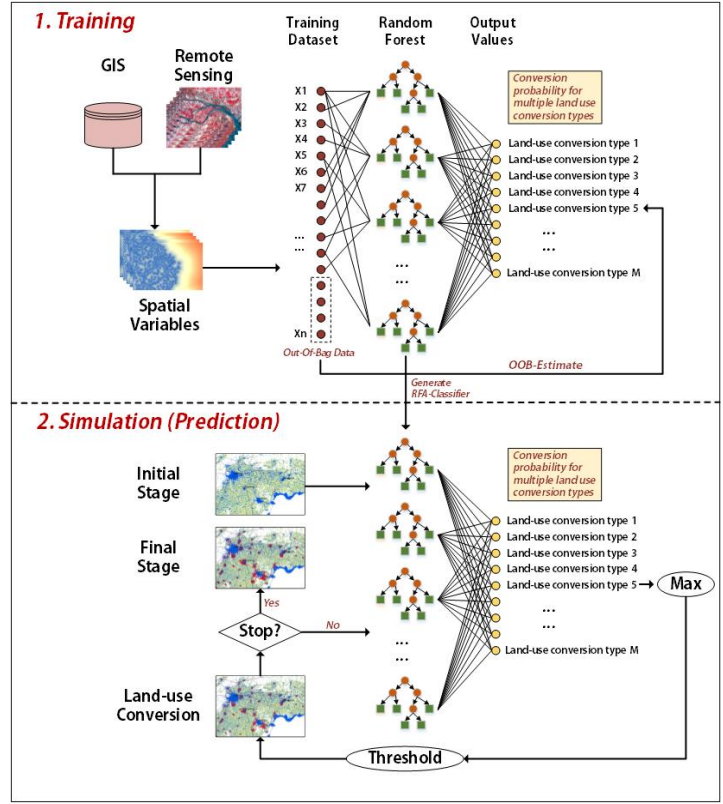

**Figure 5** structure of RFA-based CA model

The overall development probability, neighborhood effective probability, and random probability are multiplied altogether in CA model to obtain the urban conversion probability. In equation (8), $P(k,t)$ is the urban conversion probability of cell $k$ at time $t$, $Pg(k,t)$ is the overall urban development probability of cell $k$ at time $t$, $\Omega(k,t)$ is the neighborhood effective probability of cell $k$ at time $t$, and $RA$ is a random probability.

$$P(k,t) = Pg(k,t) \cdot \Omega(k,t) \cdot RA \qquad (8)$$

The overall development probability is calculated by constructing the combination classifiers of RFA. We randomly select $N$ points from land conversion data to build a training dataset $D$ of equation (9) after data being pre-processed. $S$ is the total number of spatial variables, while $Y_i$ is the $i$-th land use conversion type, $i = 1, 2, ..., 9$.

$$D = \begin{bmatrix} x_{1,1}, x_{1,2}, ..., x_{1,S}, Y_i \\ x_{2,1}, x_{2,2}, ..., x_{2,S}, Y_i \\ ... \\ x_{N,1}, x_{N,2}, ..., x_{N,S}, Y_i \end{bmatrix}$$ (9)

The land use conversion types are unevenly distributed in the land conversion dataset. So when we collect several random point sets with the same number in each land conversion type, the dataset would be repeatedly swept, that left the collecting process redundant and inefficient. Moreover, when directly collecting random points from the entire dataset, the proportion of some important land use conversion types will become very small and thus result in over-fitting problem during the training process. Therefore, we used a random sampling method to improve efficiency and to balance the quantity of samples in different conversion types. The training dataset is built as follows:

The number of random points in each line is determined by the size of the land conversion data: $pts\_per\_line = N/line\_count$.

The percentage of each land use conversion type in each line is calculated as $p_i = Num(h_k(x) = Y_i)/samples\_per\_line$, $i = 1, 2, ..., 9$ if $p_i \geq \phi$, $p_i = \phi$, and $p_j = p_j + (1-\phi)/8$ ($j \neq i$, $\phi$ is the threshold of the maximum percentage of sampling).

The number of random points that are selected from each land use conversion type is calculated as $n_i = p_i \cdot pts\_per\_line$. These randomly selected points are then joined into the training dataset $D$.

The method randomly selects $N/3$ cases from above training dataset $D$ and will repeats the random sampling $M$ times for building the combination classification model $\{h_i(X, \Theta_i), i = 1, 2...M\}$ of RFA. According to our land use data, the probability of unclassified data $\theta = [x_{\theta,1}, x_{\theta,2}, ..., x_{\theta,S}]^T$ that falls into each land conversion type is computed as follows:

$$P_i(\theta) = \frac{I(h_i(\theta) = Y_i)}{M}, i = 1, 2, ..., 9 \tag{10}$$

And the generalization error is specified mathematically as follows:

$$PE^* = P_{X,Y}(ave_k[I(h_k(X) = Y_k)] - \max_{j \neq k} ave_k[I(h_k(X) = Y_j)] < 0) \tag{11}$$

Where $P_i(x)$ is the probability of unclassified data $\theta$ that are predicted as the $i$-th land using conversion type. The overall urban conversion probability of cell $k$ at time $t$ can be computed as follows:

$$Pg(k,t) = \begin{cases} \frac{I(h_k(\theta) = Y_{nonurban-urban})}{M} & Cell(k,t) = non-urban\_area \\ 1 & Cell(k,t) = urban\_area \\ \frac{I(h_k(\theta) = Y_{limited-urban})}{M} & Cell(k,t) = limited\_development\_area \end{cases} \tag{12}$$

Whether a non-urban cell $k$ converts into an urban one is affected by its neighbor. It is obviously that when the surrounding neighbors of cell $k$ are urban area, this cell will be strongly influenced to be converted into urban land. Equations (13) and (14) describe the neighbor effective probability $\Omega(k,t)$ of cell $k$ that is located at $(x,y)$ in the dataset at time $t$. The $w$ denotes the supposed length for the window used to obtain window cells.

$$\Omega(k,t) = \frac{\sum_{i=x-w/2}^{x+w/2} \sum_{j=x-w/2}^{x+w/2} p_{Cell(i,j,t)}}{w^2} \tag{13}$$

$$p_{Cell(i,j,t)} = \begin{cases} 0 & Cell(i,j) \neq urban\_area \\ 1 & Cell(i,j) = urban\_area \end{cases} \tag{14}$$

A stochastic disturbance term is usually incorporated in CA simulation process to generate highly reasonable results (White and Engelen 1993). The disturbance can help simulation produce more accurate simulated results, which can be found in real urban systems (Li and Yeh

2002) .The random error term ($RA$) can be computed as follows (White and Engelen 1993):

$$RA = 1 + (-\ln \gamma)^{\alpha} \tag{15}$$

Where $\gamma$ is a uniform random variable within the range of 0 to 1 and $\alpha$ is a parameter for controlling the size of the stochastic perturbation (Li and Yeh 2002).

## 3.3 Iterative process

Proposed RFA–based CA is designed to simulate and predict urban expansion and farmland loss. When the urban development probability $P(k,t_i)$ calculated by the iterative equations (8) to (21) is larger than the given threshold $P_T$, $Cell(k,t_i)$ is expected to be converted into an urban area in the $i$-th iteration. In experiment, the pace of urban expansion can be controlled by adjusting the urban development threshold $P_T$. When $P_T$ is low, the urban expansion simulation process will becomes faster and the result will incorporate more details

Northam believed that the urbanization process followed an S-shaped curve and urban expansion would become stagnant after continually developing a certain amount of time (CHEN and ZHOU 2005, Northam 1979). Therefore, according to Northam's theory, we can set the stop time of the RFA based CA model when the urban expansion acceleration reaches a certain threshold of number of urban cells. Assuming that the total number of urban cells at time $t$ is $N_{UE}(t)$, the urban expansion rate $a(t)$ at time $t$ can be calculated as below:

$$a(t) = [N_{UE}(t) - N_{UE}(t-1)]/N_{UE}(t-1) \tag{16}$$

Besides, the RFA-based CA simulation also needs to meet the restriction from the total number of urban areas and the maximum iteration time. The CA iterations will not stop if none of the above three conditions has been satisfied. We design three fixed thresholds to control the convergence condition, which are minimum urban expansion acceleration $a_{given}$, minimum total number of urban expansion cells $N_{given}$, and maximum iteration times $I_{given}$ respectively. The stopping conditions of these parameters are listed as below: for urban expansion acceleration $a(t) < a_{given}$, for total number of urban expansion $N_{UE}(t) < N_{given}$, and for iteration times $I \geq I_{given}$. As soon as one of the above condition is satisfied, the RFA-based CA simulation will stop running and produce the final result.

## 3.4 Realization and parameter setting

Our research team realized all algorithms and models proposed in Section 3. Region agglomeration was realized on SPSS 9.0, and RFA algorithm was implemented by using C++ with Alglib 2.9 (http://www.alglib.net/) packages. Parallel RFA-based CA model was implemented on Kylin Linux platform (x64) of Tianhe-1 supercomputer, which is located at the National Supercomputing Center in Sun Yat-sen University, Guangzhou, China. Several open source libraries, such as open-MPI (http://www.open-mpi.org/), GDAL (http://www.gdal.org/), openMP (http://www.openmp.org) and libxml2 (http://xmlsoft.org/), were also applied to our research applications for simulation.

In the section regional agglomeration, the minimum-maximum normalization method and principal component analysis were used for extracting the principal components of socio-economic indexes. The top two principal components we chose were reserving 80.04% information by conducting PCA, and in the SPSS clustering step, we chose the centroid clustering method by setting the clustering level to 6 and adopted Pearson correlation for metrics. At last, based on the optimal clustering results, the total 34 provincial-level administrator regions were divided into different groups, which yielded 19 homogeneous economic development regions in the end.

In the urban conversion rule mining process, we implemented RFA by setting the number of decision trees 80 and the percent of training dataset and out-of-bag dataset were set to 0.60 and 0.40 respectively. The overall conversion probability from non-urban areas to urban areas was obtained by RFA-based CA model with the average OOB error of 4.85%.

During the simulation processes of RFA-based CA models, the iteration number was set to 100, and total urban growth amount was set to the same as the total changed urban area from 2000 to 2010. According to official statistical data from 2000 to 2015, Markov chains were adopted to predict growing numbers of urban cells (Arsanjani *et al.* 2013). Besides, we hired the urban expansion simulation results of ANN-based CA models to make comparison with our simulation results. The ANN-based CA models were implemented on the same platform as RFA-based CA models. Because of the random parameters of CA models, the simulation processes of all models were repeated 10 times to reduce uncertainties, and its averages were considered to make accurate evaluations. Moreover, we selected a statistical measurement named "figure of merit" (FoM) to quantify the performance of each CA model.

# 4 Results and Discussion

## 4.1 Regional agglomeration

Figure 6 and Table 1 are the results generated by regional agglomeration. As shown in Figure 6, all provinces and cities were eventually divided into 19 groups under the criterion of similarity degree in economic development status. Table 1 indicates that the results of regional agglomeration (homogeneous economic development regions) finely reflect the distribution of economic regions in China: Bohai Sea Economic Zone (Region 3), Guanzhong Economic Zone (Region 6), Chengdu-Chongqing Economic Zone (Region 9), Yangtze River Delta Economic Zone (Region 11) and Central China of five provinces (Region 10). In addition, municipality cities located within the economic regions are significantly different to the neighboring provinces in development patterns, and our result also finely reflects this feature by dividing them into different regions (Region 4,5,12,14,etc.).

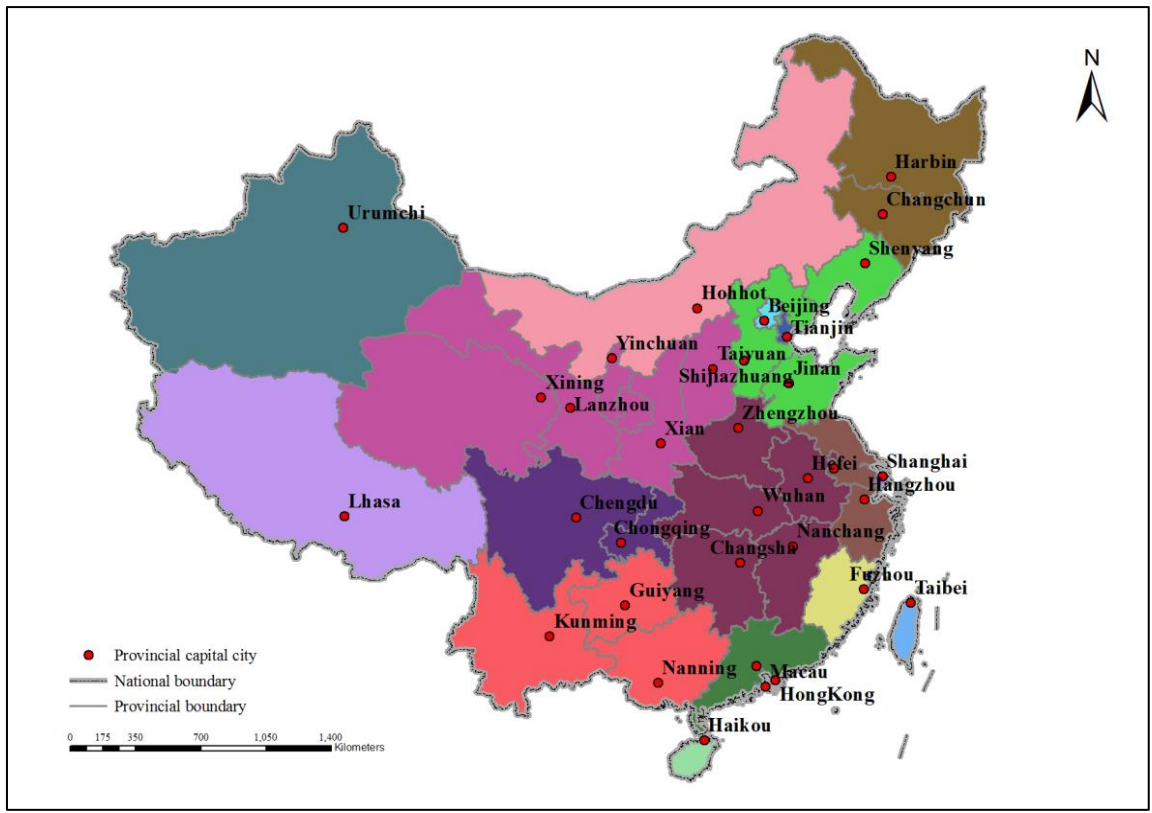

**Figure 6** homogeneous economic development regions (provinces in same color) extracted by regional clustering

| Region ID | Provinces |
|---|---|
| 1 | Heilongjiang, Jilin |
| 2 | Inner Mongolia |
| 3 | Liaoning, Hebei, Shandong |
| 4 | Beijing |
| 5 | Tianjin |
| 6 | Shanxi, Shaanxi, Gansu, Qinghai, Ningxia |
| 7 | Xinjiang |
| 8 | Tibet |
| 9 | Sichuan, Chongqing |
| 10 | Henan, Hubei, Hunan, Jiangxi, Anhui |
| 11 | Jiangsu, Zhejiang |
| 12 | Shanghai |
| 13 | Fujian |
| 14 | Guangdong |
| 15 | Yunnan, Guizhou, Guangxi |
| 16 | Taiwan |
| 17 | Hainan |
| 18 | Hongkong |
| 19 | Macau |

**Table 1** Result of regional clustering

## 4.2 Urban conversion rules

Through the homogeneous economic development regions obtained by regional agglomeration, urban conversion rules are able to be mined through the proposed RFA-based CA models. Figure 7 and Figure 8 show each variable's contribution in the process of urban conversion rules mining, while the Figure 7 shows the overall contribution weights in national level and Figure 8 reflects each spatial variable's contribution in different homogeneous economic regions.

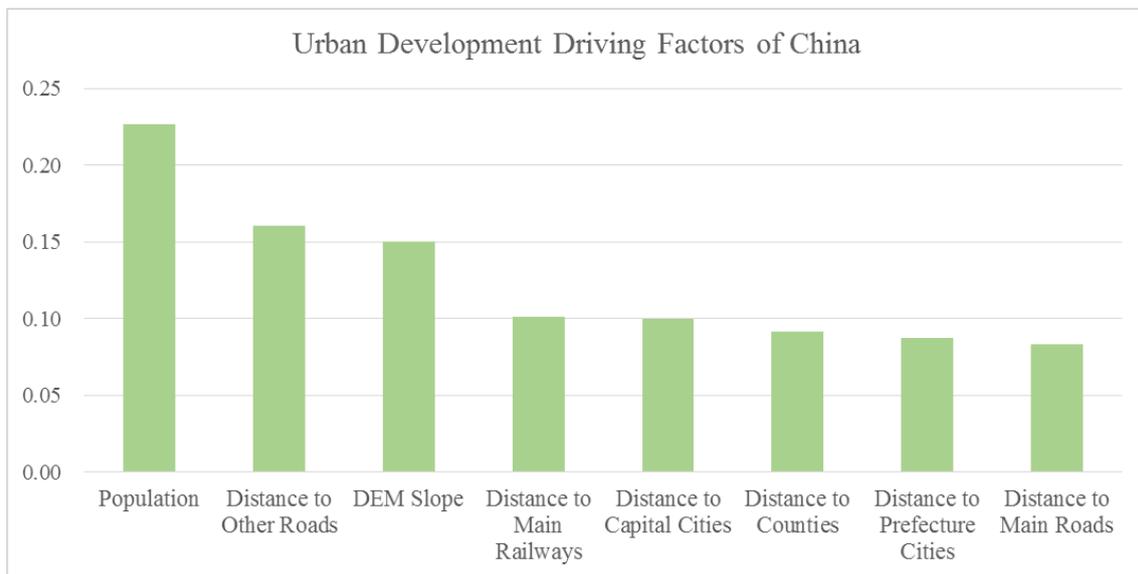

**Figure 7** Overall contribution weights of each spatial variable of China

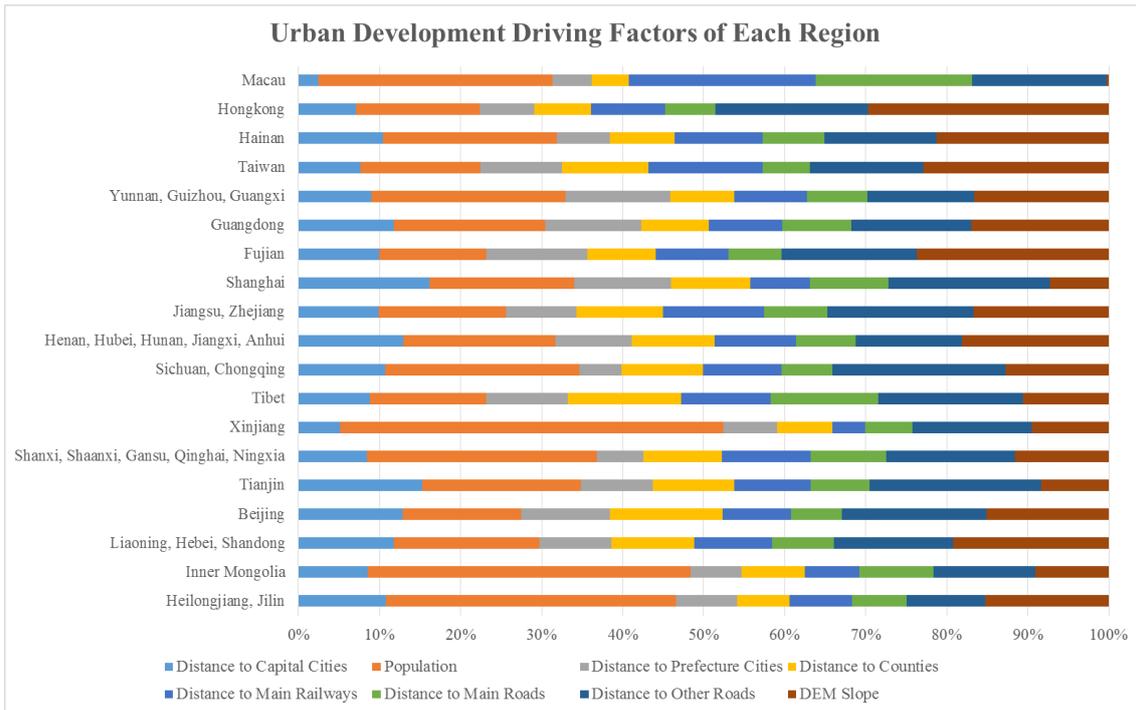

**Figure 8** Contribution of each spatial variable in different regions

Seen from a national level, the top three influence indexes of urban conversion rules are population density, roads density and terrain fluctuation. The ranking of population density helps to explain why eastern China got larger development potential than western China, and roads density can provide an explanation about the low urban development potential of Inner Mongolia, while the factor of DEM slope indicated that terrain fluctuation plays a more important role in southern China rather than northern in terms of urban development potential. Figure 8 demonstrates that the most important urban development influence factor of the provinces in Northern China is population density. The influential contribution of population density index reaches higher than 30% in Northeast China, Circum-Bohai Sea economic zone and Guanzhong region, and in Inner Mongolia, Xinjiang and Tibet, the population density even accounts for over 40% in all the influence factors. Among the provinces in the southwest and middle of China where the economic status is underdeveloped, like the five provinces in Central China (Hubei, Hunan, Henan, Anhui, Jiangxi) and three provinces in southwest China (Yunnan, Guizhou, Guangxi), the factor of terrain is as important as population density in urban development. However, for the economic developed provinces and municipalities, such as Beijing, Shanghai, Tianjin, Jiangsu and Zhejiang, the road density has stronger influence on urban development than population density and terrain conditions.

## 4.3 Urban Expansion simulation

Figure 9 shows the simulated urban expansion ratio result of China by every ten years from 2000 to 2030, and Figure 10 displays the simulation results of urban expansion for four selected economic zones in China. Due to the rapid growth of China's economy in the past two decades, China has entered a stage of 'unplanned urban sprawl' where the fast urban expansion rate can be easily found from the change of urban land ratio between Figure 9(a) and (b). As seen from Figure 9, the simulated urban land ratio for 2010, 2020 and 2030 all had a much more compact spatial structure than actual urban land ratios derived from CASM land cover data for 2000 and 2010. In the perspective of homogeneous economic development regions, we can observe that the eastern coast of China (Zhejiang and Fujian provinces) and the western China (Sichuan, Chongqing, Guizhou and Yunnan provinces) have the most rapid urban expansion speed, which may benefit from eastern coastal area's rapid economic development and government's "Westward development" policy. While in the North-East China, the growing population crisis and the fading transportation infrastructure (such as damage in road network system) are supposed to be the primary causes for its lagged pace of urban expansion. According to the statistics of urban built-up area acquired from National Bureau of Statistics of China, these observed development feature tallies with the official survey nicely.

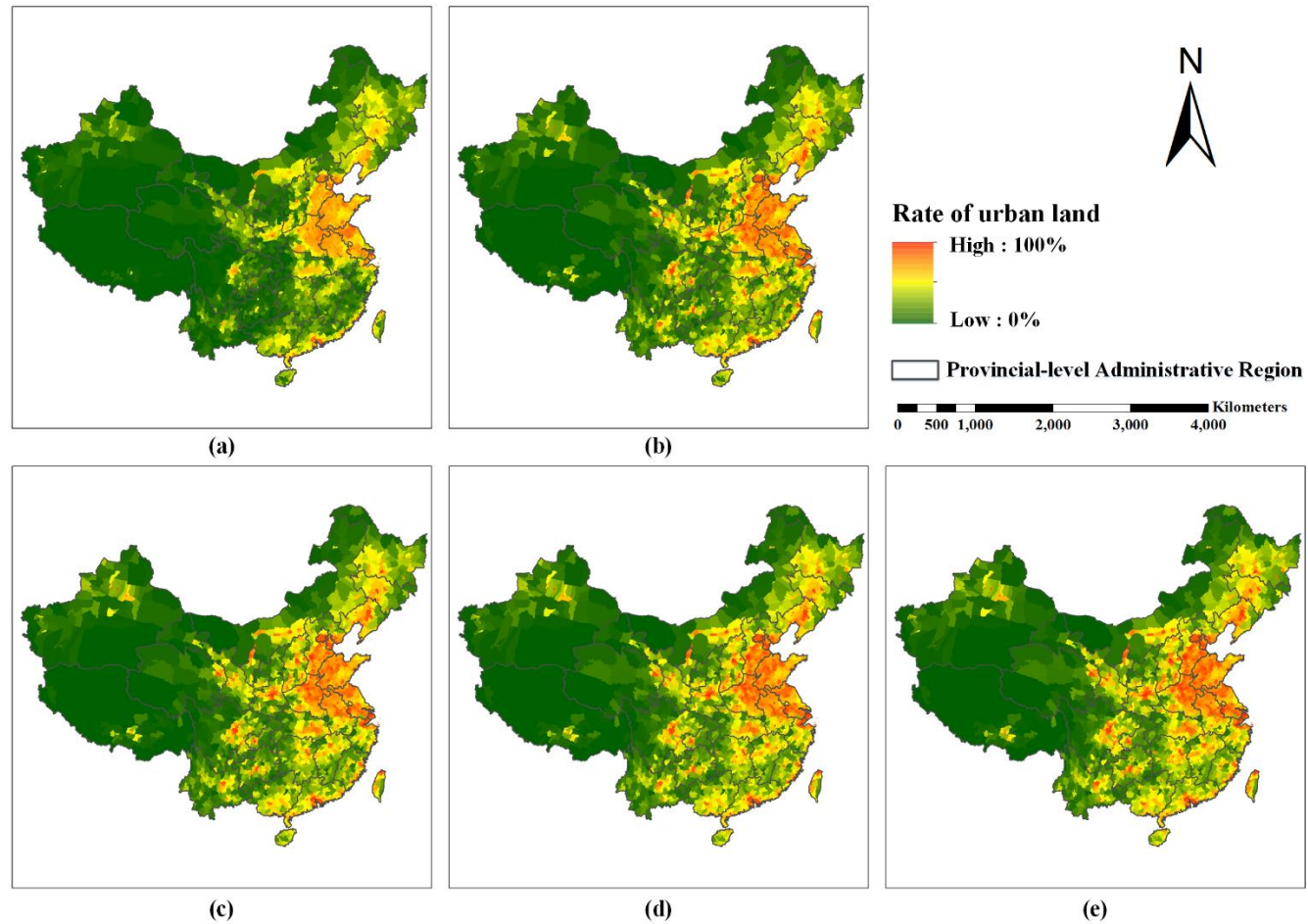

**Figure 9** Actual and simulated urban land ratios (normalization method: geometrical interval) at county level in China from 2000 to 2030 (Dynamic GIF: http://geosimulation.cn/pic/urban_sim.gif). (a) Actual rate of urban land in 2000, (b) Actual rate of urban land rate in 2010, (c) Simulated rate of urban land in 2010, (d) Simulated rate of urban land in 2020, (e) Simulated rate of urban land in 2030 .

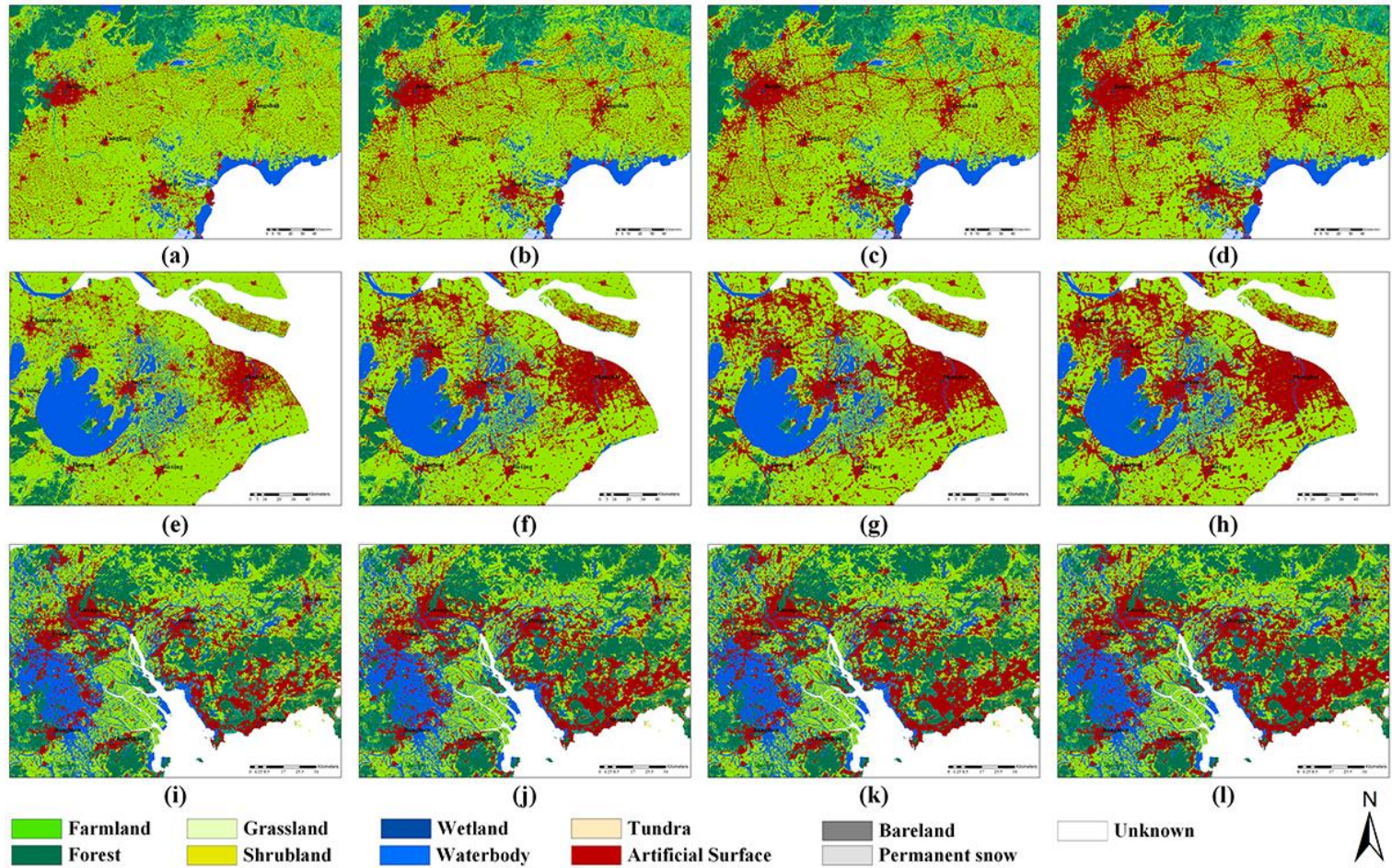

**Figure 10** Simulated results of urban expansion processes in Jing-Jin-Tang economic zone, Yangtze River economic zone and Pearl River economic zone from 2000 to 2030. (a)~(d) Simulated land cover in Jing-Jin-Tang economic zone in 2000, 2010, 2020 and 2030; (e)~(h) Simulated land cover in Yangtze River economic zone in 2000, 2010, 2020 and 2030; (i)~(l) Simulated land cover in Pearl River economic zone in 2000, 2010, 2020 and 2030.

## 4.4 Accuracy assessment

The simulation results also demonstrate that in some cities, the urban area would continue to increase while the growth speed would gradually decrease year by year. From 2000 and 2010, China entered a stage of rapid urbanization that is no parallel in history, where the ratio of urban area had increased by 65% in this decade; from 2010 onwards, the urban growth scale began to shrink; by 2030, the pace of urbanization process is expected to reduce and stabilize. Besides, urban area has the trend of agglomerating into three mega urban groups with the centers located in Beijing, Shanghai and Guangzhou. Table 2 shows the quantitative accuracy evaluation of FoM between the RFA-based and ANN-based CA models. It is noteworthy that our proposed RFA-based CA model has more correct urban expansion areas and obtain more accurate results than ANN-based CA model. The FoM value for the ANN-based model is 4.82% at the national scale, while FoM value of RFA-based model reaches up to 8.23% with 70.75% improvement. In some regions, such as Beijing, Tianjin, Shanghai, Sichuan and Chongqing, the FoM values of proposed model almost reach 20%. And the simulation results of Guangdong province obtain similar accuracy as the previous study on provincial-level urban expansion simulation (Lin and Li 2015).

The FoM values of two models seem a little bit low. Previous studies have pointed out that while CA models are applied to large scale regions or regions with small amount of observed change, FoM values tend to be relatively low(Lin and Li 2015, Pontius *et al.* 2008). In our case, the simulation scale is rather large, but the observed urban change during the period is only 0.42%. Therefore, in this case, the FoM values of this study are still acceptable, compared with similar studies (Du *et al.* 2012, Lin and Li 2015). Through the comparison between RFA-CA model and ANN-CA model, the FoM values indicate that the RFA-based CA model can perform well on simulating the complex urban conversion under a large-scale.

| Regions | Provinces | RFA-Based CA Models | | | ANN-Based CA Models | | |
|---|---|---|---|---|---|---|---|
| | | FoM | Producer's Accuracy | User's Accuracy | FoM | Producer's Accuracy | User's Accuracy |
| 1 | Heilongjiang, Jilin | 3.22% | 5.07% | 8.11% | 1.01% | 1.89% | 2.12% |
| 2 | Inner Mongolia | 3.88% | 14.89% | 4.98% | 2.25% | 7.62% | 3.10% |
| 3 | Liaoning, Hebei, Shandong | 7.61% | 12.11% | 16.98% | 5.62% | 12.86% | 9.09% |
| 4 | Beijing | 15.58% | 32.55% | 23.00% | 8.00% | 39.51% | 9.11% |
| 5 | Tianjin | 16.59% | 33.33% | 24.82% | 5.86% | 31.43% | 6.72% |
| 6 | Shanxi, Shaanxi, Gansu, Qinghai, Ningxia | 6.88% | 9.41% | 20.35% | 4.28% | 6.37% | 11.50% |
| 7 | Xinjiang | 7.48% | 25.03% | 9.64% | 2.48% | 6.06% | 4.04% |
| 8 | Tibet | 5.33% | 13.95% | 7.94% | 1.06% | 1.27% | 6.12% |
| 9 | Sichuan, Chongqing | 17.10% | 20.53% | 50.57% | 7.46% | 10.09% | 22.23% |
| 10 | Henan, Hubei, Hunan, Jiangxi, Anhui | 8.38% | 14.12% | 17.08% | 5.38% | 10.96% | 9.57% |
| 11 | Jiangsu, Zhejiang | 10.94% | 28.27% | 15.14% | 7.92% | 30.49% | 9.67% |
| 12 | Shanghai | 20.76% | 36.04% | 32.86% | 8.38% | 35.68% | 9.87% |
| 13 | Fujian | 5.86% | 16.97% | 8.22% | 6.58% | 22.08% | 8.57% |
| 14 | Guangdong | 9.35% | 28.77% | 12.17% | 5.05% | 19.22% | 6.41% |
| 15 | Yunnan, Guizhou, Guangxi | 4.64% | 5.05% | 36.06% | 2.81% | 3.09% | 23.73% |
| 16 | Taiwan | 7.29% | 15.33% | 12.20% | 5.45% | 23.68% | 6.62% |
| 17 | Hainan | 5.68% | 7.34% | 20.10% | 5.74% | 10.24% | 11.54% |
| 18 | Hongkong | 3.11% | 3.23% | 45.45% | 5.26% | 6.12% | 27.27% |
| 19 | Macau | 32.78% | 72.28% | 37.52% | 28.61% | 64.38% | 33.99% |
| | **National level** | **8.23%** | **14.50%** | **15.99%** | **4.82%** | **10.08%** | **8.44%** |

**Table 2** Comparison of urban simulation results between RFA-based CA models and ANN-based CA models (FoM)

## 4.5 Farmland loss simulation

Rapid urbanization in China was at the expense of tremendous amounts of farmland, which accounts for most of the farmland loss in China (Ding 2003, Tan *et al.* 2005). Therefore, the predictive urban expansion results of proposed RFA-based CA model can be used to simulate the overall farmland loss. Figure 11 displays the simulated farmland proportions at national level from 2000 to 2030. We can observe from the simulation result that during the period of 2000-2010, the erosion amount of farmland caused by the rapid urban expansion accounts for more than half of the total erosion amount for the overall 30 years from 2000 to 2030. The situation of farmland erosion is even much more serious in the developed eastern coast of China. The simulated farmland loss results indicate that the total farmland of China will decrease 5.72% from 2000 to 2030 due to the rapid urbanization process, and the annual average loss of farmland is about 2,440 km$^2$ per year. The great loss of farmland poses critical challenge to agricultural security and food security.

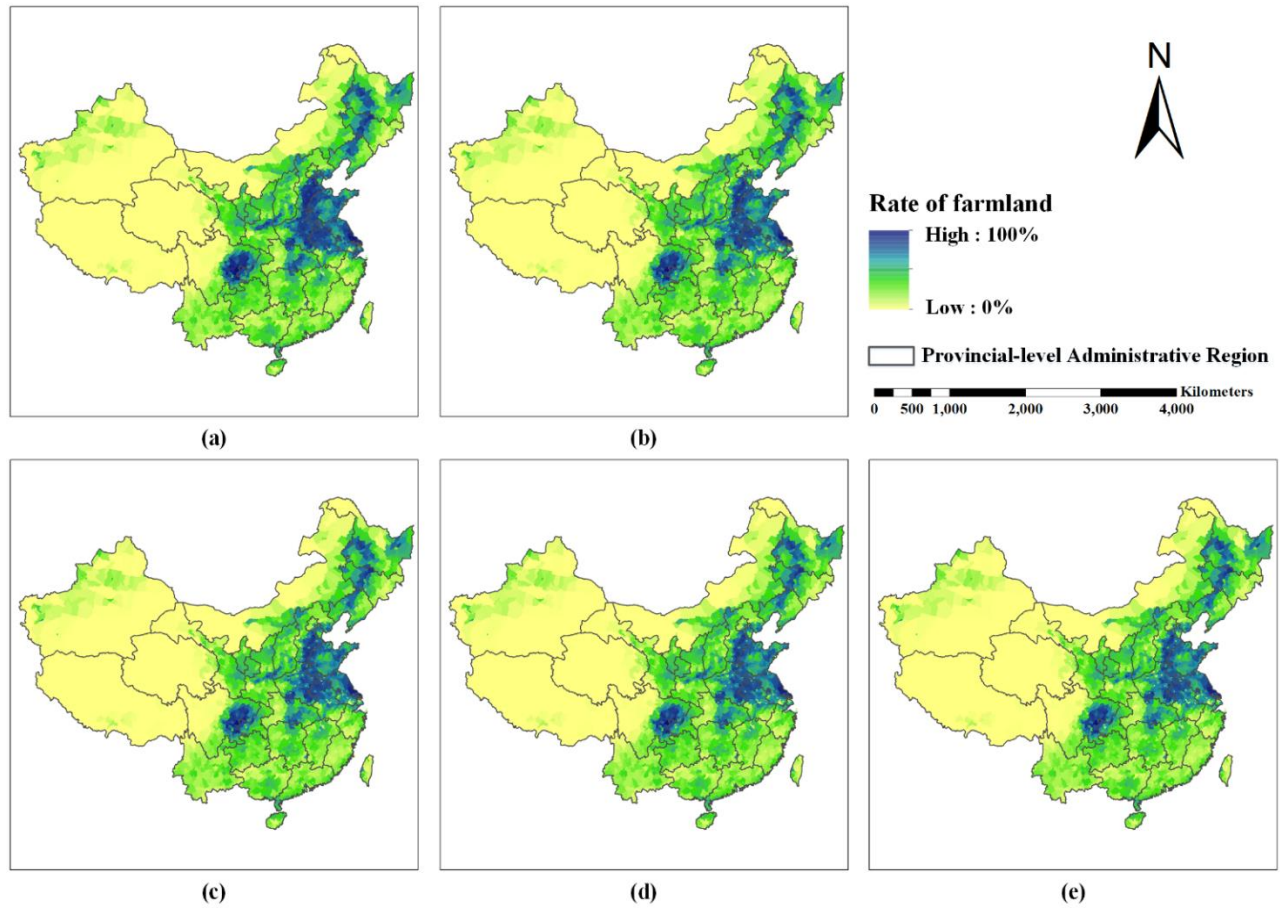

**Figure 11** Actual farmland and simulated farmland loss ratio at national level in China from 2000 to 2030 (Dynamic GIF: http://geosimulation.cn/pic/farmland_sim.gif). (a) Actual rate of farmland in 2000, (b) Actual rate of farmland rate in 2010, (c) Simulated rate of farmland in 2010, (d) Simulated rate of farmland in 2020, (e) Simulated rate of farmland in 2030

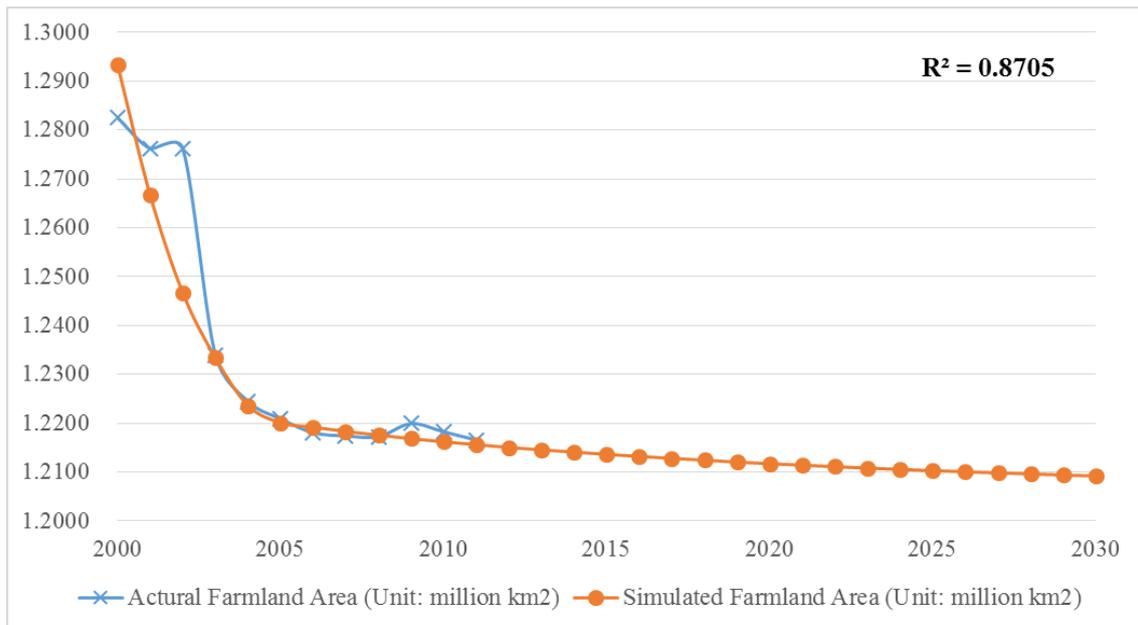

**Figure 12** Comparison between simulated farmland area and actual farmland area. Yellow line shows the simulation and prediction results of farmland area change from 2000 to 2030, and blue line shows the actual farmland area change from 2000 to 2011.

In the national second land use survey, the government used a new statistical method to compute the farmland area, which causes the total farmland area in 2012 to be larger than that of the previous land surveys (Xinhuanet 2013). Thus, the statistical arable data of the years of 2012 and 2013 were not used for accuracy assessment in our study. By comparing the simulation results of farmland area with the official statistical data from 2000 to 2011, we obtained the results shown in Figure 12. Figure 12 demonstrates that the simulation results of farmland area are close to the reality where the standard deviation between the simulations of farmland area and the actual farmland areas is 0.02, and the R-squared $R^2$ is 0.87. The accuracy results indicate that the rapid urban expansion process is the most important factor of farmland loss in China compared with the other factors, such as land degradation, natural hazard, climatic variation and ecological change. Compared with the year 2000, 1.28 million km² farmland areas have been rapidly reduced to less than 1.22 million km² in 2005, and its annual reduction rate is as fast as 0.07% in this period. It is only after 2005 that the 70  dangerously rapid loss of farmland was curbed. Since then, the annual farmland loss rate was kept restricted within 0.03%. According to the simulation results, China is able to preserve the 1.20 million km$^2$ farmland from crossing the "arable-land red line" within the next 20 years, but the situation remains severe.

## 5 Conclusions

75  In this study, in order to address the key problems existed in multi-rules digging of urban development and data mining under the scenario of national-scale at high spatial resolution, firstly we aggregated the provincial-level administrator zones to several homogeneous economic development regions by regional agglomeration. Then, a RFA based CA model was carried out to obtain urban conversion rules and different spatial variables' contribution of
80  urban development in multi-scale. Finally, proposed RFA-based CA models were used to simulate urban expansion and farmland loss processes from 2000 to 2030 in a fine scale of 30 meters. Compared with the simulation results of ANN based CA models, the accuracy of proposed RFA-CA models improved about 70.75%. This study effectively illustrates that the essential driver of farmland loss is the rapid urbanization of China. Furthermore, the simulation
85  results indicate that in the next 20 years, the farmland loss rate is expected to slow down gradually, and the farmland area will stabilize. However, China is still in the stage of rapid urban expansion, and considering its high relevance with farmland loss, the future of farmland preservation still remains severe.

Additionally, this study provides a large-scale analytical framework of RFA-based CA
90  model for complicated geographical phenomena simulation at high spatial resolution, and can be utilized at larger scale like global scope in the future. The performance of our proposed RFA-based CA model is better than previous CA model for large-scale simulation, and the accuracy of RFA-CA model can still be improved under our proposed framework. In future studies, System Dynamics and more practical factors will be introduced in patch-based CA models to control the microsimulation process and improve the simulation performance based on current work.

## Acknowledgment

This study was supported by the National Natural Science Foundation of China (Grant No. 41671398), the Key National Natural Science Foundation of China (Grant No. 41531176) and the National Natural Science Foundation of China (Grant No. 41601420).